\documentclass[%
 aip,
cp,  
 amsmath,amssymb,
 reprint,%
]{revtex4-2}

\usepackage{graphicx}
\usepackage{dcolumn}
\usepackage{bm}
\usepackage[dvipsnames]{xcolor}
\usepackage[utf8]{inputenc}
\usepackage[T1]{fontenc}
\usepackage{mathptmx} 
\usepackage{rotating}
\usepackage{float}
\usepackage{}
\begin{document}
{\color{gray}\centerline{Manuscript submitted to the Journal of Undergraduate Research in Physics.}}
\title{Modeling Supernovae as an Optically Thick Fireball}

\author{Jacob Marshall} 
 \email{jacob.marshall.astr@gmail.com}
\author{Scott Severson}%
 \email{scott.severson@sonoma.edu}
\affiliation{
  Sonoma State University Department of Physics and Astronomy. 1801 E. Cotati Avenue, Rohnert Park, California.
}


\date{\today} 

\begin{abstract}

We calculate the properties of 135 stellar supernovae using data from the Open Supernova Catalog. We generate temperatures, radii, luminosities, and expansion velocities using a spherically symmetric optically thick fireball model. These modeled parameters reveal trends that are common across different types of supernovae. We have identified distinct phases that appear across Type Ia, II, II P, and IIb supernovae. We note that there is a long period of reasonable continuous growth (Phase 1), giving credence to our simple model of an optically thick fireball. The modeled radius reaches a maximum value beyond which it is flat or decreases (Phase 2). The temperature we observe at the maximum modeled radius, 4500 K, suggests that the loss of opacity due to electron recombination sets the timeline where our optically thick model no longer applies. We observe the fastest modeled fireball velocities, largest modeled fireball radii, and maximum modeled luminosities for Type Ia supernovae. As a group, Type Ia supernovae reach a maximum luminosity that is 8.5 times more luminous than Type II supernovae. We present a summary table that contains modeled parameters of supernovae and their timings by supernova classification type.

\end{abstract}
\maketitle

\section{Introduction}
We apply a physically motivated yet simple model to the photometric data within the Open Supernova Catalog\cite{2017ApJ...835...64G}. Treating the emitted light as originating from an optically thick sphere we model their progression by identifying trends in radial growth, temperature, and luminosity over time. These descriptions are accompanied by figures and tables generated using the data from the catalog, highlighting key moments during the lifetime of the event. The sample modeled in this paper consists of 135 supernovae. These supernovae are composed of different types with 86 being Type Ia, 19 Type II, 19 Type II P, and 11 Type IIb, as discussed below.

\subsection{Supernovae}
In this section, we provide a brief overview of the types of supernovae we model from the Open Supernova Catalog. A supernova is an event that occurs as a star reaches the end of its life. Supernovae are some of the most energetic and luminous events in the universe and have the potential to be more luminous than their host galaxies. Our sample includes Type Ia, II, II P, and IIb supernovae.\par
 
A Type Ia supernova occurs in a binary system consisting a white dwarf and a companion star. A white dwarf is an end state of the core of a low-mass star that can no longer produce fusion. As the star runs out of its nuclear fuel it no longer produces pressure to counteract its own gravity. The star will begin to collapse until the electrons reach densities where electron degeneracy begins to play a role. Electron degeneracy pressure allows the white dwarf to prevent gravitational collapse, but it no longer produces energy through fusion. Since the atoms within the star are so close, it is extremely dense. \cite{Hillebrandt_2001}  \par 

The nature of electron degeneracy sets a maximum mass for an object on this size scale. This mass is around 1.44 solar masses and is known as the Chandrasekhar limit. If a white dwarf exceeds this mass, the result is a Type Ia supernova. A white dwarf can exceed this mass by accreting matter from its partner in a binary system. As the stars orbit each other, gravitational interactions cause matter from the partner star to settle on the surface of the white dwarf. When the white dwarf mass surpasses the Chandrasekhar limit, it is possible that small reactions on the surface involving this accreted matter will cause thermonuclear fusion that lead to the runaway explosion of the white dwarf and thus a supernova.  \par 

Type II supernovae are different from the Type Ia. A star that is between 10 and 50 solar masses will fuse elements heavier than hydrogen and helium in its core. As the star fuses heavier and heavier elements, these interactions produces less energy. As the star gets older and hotter it will begin to form an iron core, which cannot produce energy via fusion and thus cannot counteract the force of gravity. As the core becomes inert, the weight of the star's outer layers compress the core until even electron degeneracy pressure cannot prevent gravitational collapse. As the core implodes, the outer layers collapse and it is energetic enough to produce neutrons via neutronization where protons and electrons fuse to create a neutron and an electron neutrino. The collapse continues until the neutrons in the new core are unable to be compressed further. When this happens, a shock wave ripples through the remaining matter sending it outward fast enough to escape the gravity of the core. The ejection of this matter is referred to as a Type II supernova. These supernovae are identified using the hydrogen in their spectra as they have outer layers with unprocessed hydrogen. \cite{Hiramatsu_2021}  \par 

Similar to Type II supernovae, Type IIb supernovae are also the result of a massive star exhausting its supply of energy from fusion in its core. A difference is a weaker hydrogen line in its spectra which can even disappear over time. This discrepancy in the quantity of hydrogen is attributed to a star that has let go of its outer hydrogen layers or a star in a binary system that has had its hydrogen layers stripped away by its partner through gravitational interactions. A unique feature of a Type IIb supernovae is that it can have two peaks in its light curve. \cite{Sravan_2020}  \par 

Type II P supernovae are Type II supernovae that have a slower decrease in luminosity, resulting in a light curve that has a shoulder or plateau. This plateau is the result of ionization of hydrogen in the outermost envelope of the explosion. As this happens, the opacity of this envelope increases, keeping photons contained within the envelope for a longer period of time thus the plateau in luminosity. As the hydrogen envelope cools, protons recombine with their electron to form hydrogen atoms again. The outer envelope becomes transparent enough for photons to escape and we see the luminosity decrease at a greater rate. \cite{Hiramatsu_2021} 

\subsection{Catalog}

The Open Supernova Catalog is a database containing metadata, spectra, and photometric data (light curves) from across the electromagnetic spectrum including: radio, ultraviolet, infrared, and X-ray measurements. Currently, the catalog has data for more than 14,000 events. These data are collected from many repositories and publications. The purpose of creating such a meta-catalog is to make data accessible to the scientific community and public in a central repository. \cite{2017ApJ...835...64G} 

We wrote a set of scripts to download the photometric data for supernovae in the catalog. We selected supernovae that had ample measurements of B band and V  band photometric data. \cite{1953ApJ...117..313J} We use these B band (mean wavelength: 442 nm) and V band (mean wavelength: 540 nm) measurements to approximate a color temperature in a uniform way.   We parsed these approximately 400 supernovae and selected the 217 that had ample magnitude measurements for us to perform our analysis. Of these 217 supernovae, 135 had light curves that were complete, allowing us to determine the parameters of the model: estimates of luminosity, radius, and temperature.

The following sections will discuss our fireball model and how we applied it to these events. We will further discuss the results of applying this model to the data set, and what these results mean within the context of our model. We will highlight the limits of our model, showing where the assumption of an optically thick fireball no longer holds.

\section{Modeling the data}

Our model treats the supernova explosion as an optically thick fireball, where the surface of the photosphere is in thermodynamic equilibrium.  This allows us to use the color magnitudes in the B (blue) and V (visual) bands to estimate certain parameters of the fireball. To collect, parse, and evaluate the magnitude measurements for a supernova, we wrote a program that collects photometric data from the Open Supernova Catalog. Our final sample includes 135 Type Ia, II, IIb, and IIP supernovae based on counts of photometric data. The data was parsed to find the B magnitude, V magnitude, observation time, and luminosity distance. B and V band measurements were paired when they were taken no more than 72 minutes apart. The pairing ensured that the measurements were roughly contemporaneous, and the 72 minute figure (0.05 days) provided ample matching in the survey data.  These magnitude measurements and distances are used to estimate the temperature and the luminosity of the fireball. When approximating the luminosity, we converted the apparent magnitudes given in the catalog to absolute magnitudes. The magnitudes and distance are used in Equation 1, where M is the absolute magnitude, m is the apparent magnitude, and d is the distance to the supernova in parsecs. \cite{Carroll} 
\begin{eqnarray} M = m - 5log(d/10pc)
\end{eqnarray} 

This fireball can then be treated as a blackbody, allowing us to use the B and V band measurements to estimate the surface temperature via Planck's Law. A greater temperature results in a larger ratio of B to V band flux.  As a supernova cools, the B band flux decreases quicker than the V band flux. We approximated the temperature at the surface of the fireball using the following equation, known as Ballesteros' formula. \cite{2012}

\begin{eqnarray}
T = 4600(\frac{1}{0.92(B-V)+1.7} + \frac{1}{0.92(B-V)+0.62}) K
\end{eqnarray}

Since these magnitudes are logarithmic, their difference, represented by B-V, is a ratio of the fluxes in these bands. \par

To derive the total luminosity across all wavelengths, bolometric luminosity, we must apply a correction to account for flux in wavelengths outside the B and V bands. Our fireball model treats the supernova as a black body, so at a given temperature there is a known ratio of the light emitted in the V band versus the total amount of light emitted over all wavelengths. The emission at a given wavelength and temperature is governed by the Planck function, Equation \ref{pfunc}.

\begin{eqnarray}
B_\lambda(\lambda,T) = \frac{2hc^5}{\lambda^5}\frac{1}{e^{\frac{hc}{\lambda k_BT}}-1}
\label{pfunc}
\end{eqnarray} 

Our initial calculation of a supernova's total luminosity uses the Sun as a bolometric reference. Our calculation of the supernova's bolometric luminosity is shown in Equation \ref{lbol} below. We begin by scaling the Sun's bolometric luminosity ($L_{bol\odot} $) by the difference of the supernova's ($M_{VSN} $) and Sun's absolute magnitude ($M_{V\odot} $) in the V band. There are two additional terms that capture the temperature related difference in the supernovae versus the Sun's emission. The fraction of the light emitted in the V band is related to the integral of the Planck function at the temperatures of the fireball ($T_{SN}$) and the Sun ($ T_{\odot}$). The Planck function is integrated over the wavelengths of the V band with $\lambda_{min}$ = 507nm and $\lambda_{max}$ = 595nm. Finally, the Stefan-Boltzmann law shows that total blackbody flux scales as $T^4$, and we scale the bolometric luminosity as $ (\frac{T_{SN}}{T_{\odot}})^4$. 

\begin{eqnarray}
L_{bolSN}= (\frac{T_{SN}}{T_{\odot}})^4\times\frac{\int_{\lambda_{min}}^{\lambda_{max}} B_\lambda(\lambda, T_\odot)d\lambda}{\int_{\lambda_{min}}^{\lambda_{max}} B_\lambda(\lambda, T_{SN})d\lambda}\times10^{\frac{M_{VSN}-M_{V\odot}}{-2.5}} \times L_{bol\odot} W
\label{lbol}
\end{eqnarray}

To test the validity of our model, we compared our values to alternate bolometric calculations used in the field. Bolometric corrections done by Martinez et al.\cite{refId0}, Pechja and Prieto  \cite{Pejcha_2015,Pejcha_20152}, Bersten and Hamuy \cite{Bersten_2009}, and Lyman et al. \cite{Lyman_2013} are numerically calculated using spectra at all wavelengths. Equation 5 shows that in these methodologies, the total bolometric luminosity is calculated from the absolute magnitude of the supernova in the V band, a characteristic luminosity ($3.0128\times10^{28}$W), and a bolometric correction, BC. In these methodologies the bolometric correction is dependent on color indices (e.g B-V) and characterized using a polynomial fit. 

\begin{eqnarray}
L_{bol}=3.0128\times 10^{28} \times 10^{\frac{M_{VSN}+BC}{-2.5}} W 
\label{6}
\end{eqnarray}

\begin{figure}[h]

\includegraphics[width=\linewidth]{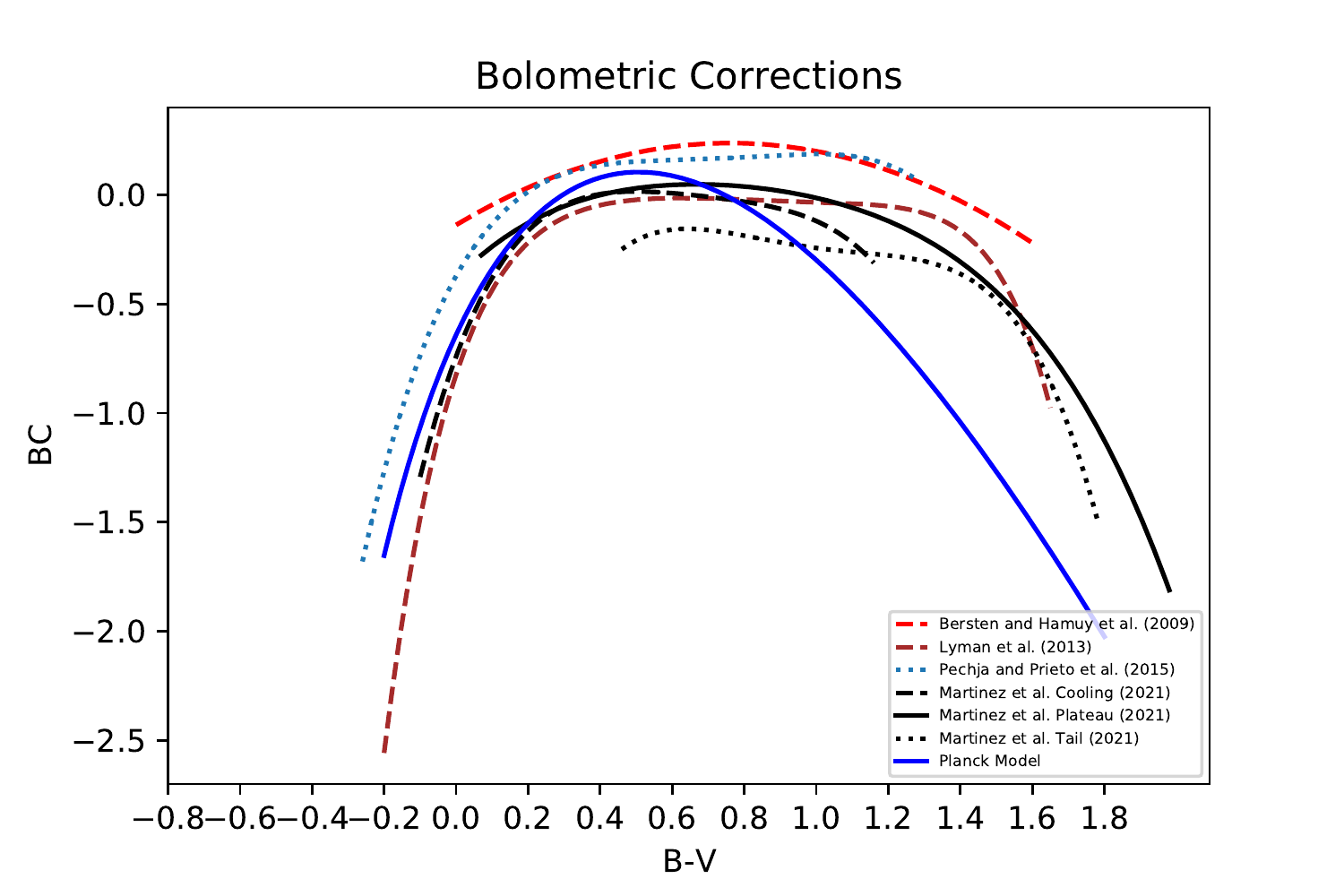}
\caption{Plot summarizing the behavior of various bolometric corrections versus B-V photometric color. We include results derived from spectra done by Martinez et al.\cite{refId0}, Pechja and Prieto \cite{Pejcha_2015,Pejcha_20152}, Bersten and Hamuy \cite{Bersten_2009}, and Lyman et al. \cite{Lyman_2013} and our own correction modeled by Planck blackbody radiation. Low B-V is bluer and hotter while larger B-V is cooler and redder emission. There is substantial negative correction while the supernova is initially hot. For mid-range values of B-V, there is a more moderate correction, often close to $BC=0$. As the value of B-V increases beyond 1.4, there is another increase in the negative correction. While our Planck model shows the same general trends as established methods, we see over-correction as B-V values increase. Our correction deviates from the literature around temperatures of 4500K (B-V = 1.1). As we show later in the paper, this is where we believe the fireball is no longer optically thick. Due to this change in optical thickness, our Planck blackbody model no longer applies. For the purpose of this paper, we have decided to adopt the Martinez Cooling and Plateau corrections as to not over-estimate the bolometric luminosity at low temperatures. We apply the Cooling correction for B-V values less than 0.6 and the Plateau correction for B-V greater than this.  }
\centering
\label{BCs}
\end{figure}

To compare our modeled bolometric luminosity to the existing literature, we adopt the form of a reverse engineered bolometric correction as a function of B-V. Figure \ref{BCs} shows the corrections from the literature and our reverse engineered model. The bolometric correction of our model  shows the same general trends as the literature. The value of the bolometric correction gets more negative at small B-V (high temperature) and large B-V (cool temperature). On the high temperature end, our model is in keeping with values from the literature, for cool temperatures (B-V $\gtrsim$ 1) our values deviate from the established bolometric corrections. Treating the supernova as a perfect blackbody and using the Planck function to estimate the bolometric correction appears to be over-estimating the luminosity at temperatures below 4500K.
In fact, when we applied our bolometric correction, we noticed this over-estimation as extended shoulders across our preliminary luminosity curves. Thus, we have adopted the bolometric corrections from Martinez et al.\cite{refId0}. Our adoption of the Martinez models utilizes the Martinez Cooling correction for B-V values less than 0.6 (5900K) and the Martinez Plateau correction for B-V values greater than 0.6. 

For the purposes of this paper we forego correcting for the effects of extinction or loss of flux in the line of site between the supernovae and the telescope. Thus, the values of our luminosity and total radiative energy are lower limits. This correction over our sample is left as future work.

The modeled fireball radius is calculated using the Stefan-Boltzmann law. The Stefan-Boltzmann Law states 
\begin{eqnarray} 
L = 4\pi R^2\sigma T^4 \label{eq:SB} 
\end{eqnarray}  

where L is the total luminosity, $4\pi R^2$ is the surface area of the photosphere, and $\sigma$ is the Stefan-Boltzmann constant which has a value of approximately $5.67 \times 10^{-8} W/m^2K^4$. Solving this equation for R give us 
\begin{eqnarray} 
R= \sqrt{L/4\pi \sigma T^4}
\end{eqnarray} 
which is used to approximate the radius of the fireball at each point in time. \cite{Carroll}\par

We created scatter plots of each event's approximated radii, temperatures, and luminosities over time. We used these plots to determine similarities and differences across the supernova types in our survey. The discussion of our observations is reserved for the next section.

\section{Analysis}

\begin{figure}[h]

\includegraphics[width=\linewidth]{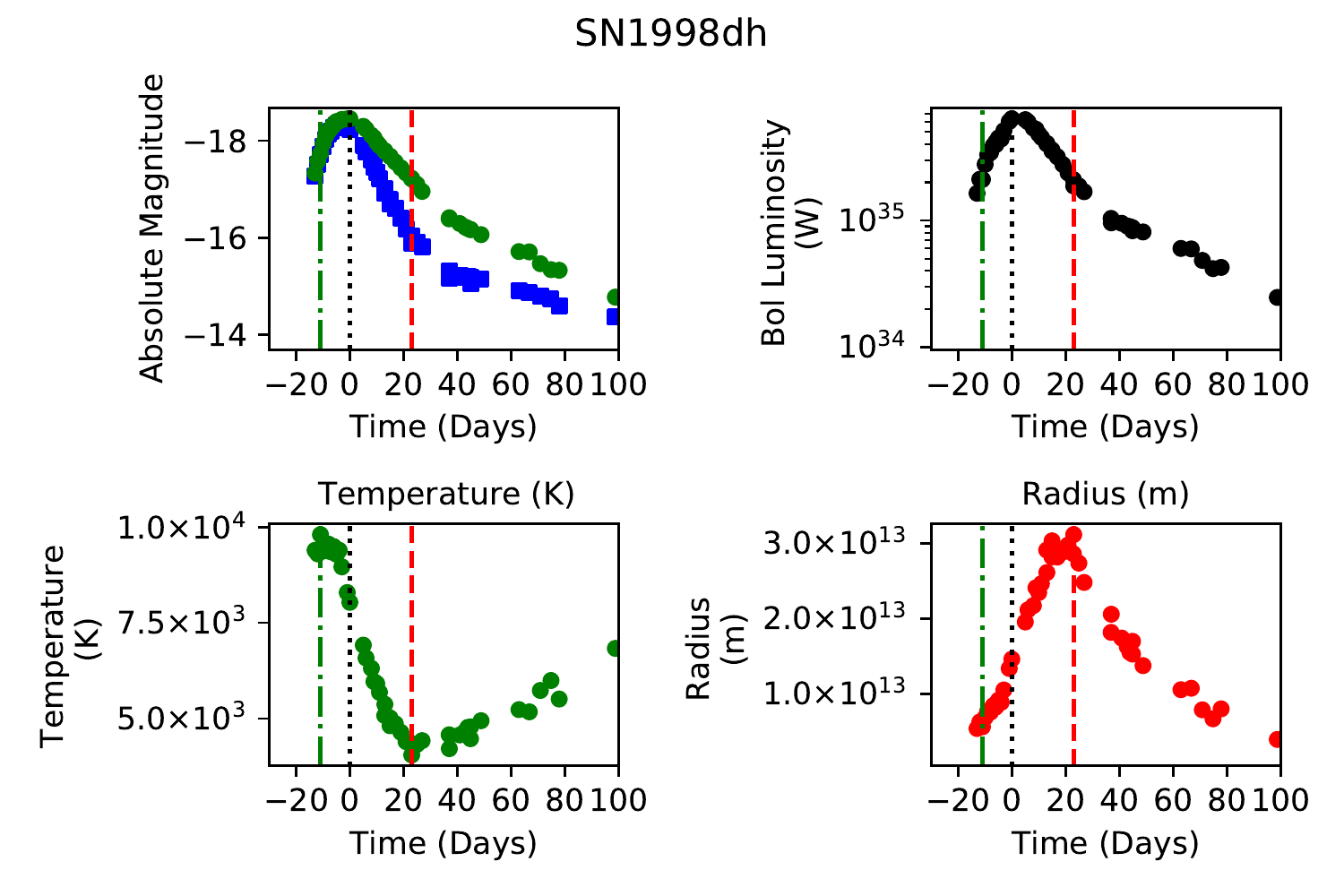}
\caption{Plots of the absolute magnitude in the B and V band, bolometric luminosity, radius, and temperature for a sample Type Ia supernova (SN1998dh) generated using photometric data from the Open Supernova Catalog. The color magnitudes and the distance to the supernovae allow us to approximate its luminosity and the difference between B and V magnitudes is an indicator of the fireball's temperature. Temperature, luminosity, and radius are related through the Stefan-Boltzmann equation (Equation \ref{eq:SB}). These plots provide a timeline of the supernova. The fireball begins by growing, heating, and getting brighter. Eventually it dims and cools while continuing to grow.  This timeline led us to create a phase model of the lifetime of the event. The dash dotted (green) grid line signifies the maximum modeled temperature, the transition from phase 1a to phase 1b. The dotted (black) grid line signifies maximum luminosity, the transition from phase 1b to 1c. The dashed (red) grid line signifies maximum radius, the transition from phase 1c to 2. Beyond phase 2, we see the modeled radius begin to decrease, evidence of a change in optical thickness signifying the model assumptions no longer hold.}
\centering
\label{4SQ}
\end{figure}

The large and comprehensive sample allowed us to identify behaviors that are consistent and those that vary across the supernova types in our survey. Figure \ref{4SQ} contains example plots for each modeled parameter: B and V band absolute magnitude, bolometric luminosity, temperature, and radius for Type Ia supernova SN1998dh. After the onset of the supernova, we see an increase in the values of luminosity, temperature, and radius. In this figure, shortly after the onset the temperature reaches its maximum value, followed by the maximum luminosity. The modeled fireball continues to show radial growth for about another three weeks, reaching a maximum value at the time of an inflection in temperature as well as B and V band luminosity. The general timings of the maxima and behavior of the parameters over the supernova duration is a feature common to all supernovae in our analysis. \par

The modeled radial growth of all supernovae types indicate a fireball that grows in size at early times until well after the peak luminosity. At some point the modeled radius eventually breaks this growth trend. This matched our original hypothesis, that while the fireball was optically thick, we would observe  growth, and that as the fireball became optically thin, the modeled "radius" would show a flattening or decrease as the model assumptions are no longer valid. To be clear, the outward expansion of the supernova itself is not halting, instead we believe this decreasing modeled radius is an indicator of a transition from an optically thick to optically thin fireball. An example physical mechanism for this transition is the recombination of ionized hydrogen to neutral hydrogen and the concomitant loss of the electron scattering opacity source. This transition in optical thickness is apparent in each supernova in the sample. Another indicator of this transition is the increase in modeled temperature around the same time as the turnover in the radius, which is common across all types, perhaps an indication that we are looking at higher temperature inner portions of the fireball as it becomes optically thin. \par

Of our initial sample of 217 supernovae, there were 82 from all types that we deemed unfit for inclusion in our statistical analysis. These supernovae contained erroneous or incomplete light curves, light curves with multiple conflicting measurements taken at the same time, or non-physical values. These are not included  in the statistical analysis in Table \ref{finsum}.

The following sections will discuss in more detail how we came to identify the phases of the modeled supernova behaviour as well as analyzing differences and similarities across different supernovae types.

\subsection{Phases}
Observing the behavior of the supernova over its duration shows trends that are common among the different types. We observe a  rise in the temperature, radius, and luminosity until the peak temperature is achieved. As the event continues, the photosphere continues to expand and a maximum luminosity is reached. After this point, the photosphere continues to grow as it continues to cool and becomes less luminous. We see a point where the estimated radius reaches a maximum. At this time we observe the temperature increase again, likely marking the point where we are no longer looking at the surface of the photosphere. This is likely the result of a change in optical thickness. Our model assumes the photosphere is optically thick. We have used this information to create four distinct phases of the event over its duration. The timing of these phases and the paired values of maximum temperature, luminosity, and radius are summarized in  Table \ref{finsum}. \par

We have named the period of time where the supernova is optically thick Phase 1. Phase 1 is broken up in to sub phases based on our modeled luminosities, temperatures and radii. Phase 1a begins immediately upon the supernova onset and sees the rise of temperature, radius, and luminosity. The onset time is not always well-determined due to the nature of the underlying observations and when they first detect the supernova. For Table \ref{finsum} we include an estimated onset time for Supernovae where we have an early-time measured luminosity that is is 10\% or less than the maximum luminosity. This phase sees the fireball becoming more luminous, expanding, and heating  until the maximum temperature is reached.\par

The next phase, Phase 1b, begins at this time of maximum temperature. Throughout Phase 1b the supernova radius and luminosity are increasing while the temperature is decreasing. The continued brightening of the fireball while the temperature is decreasing is due to the growth of the photosphere outweighing the temperature reduction and is governed by the Stefan-Boltzmann Law given in Equation \ref{eq:SB}. The end of Phase 1b occurs when the supernova reaches its peak luminosity.

Phase 1c begins with the time of peak luminosity. This is defined to be the time zero as is the custom in the field. Timings given in our tables and plots are therefore referenced to the time of peak luminosity, with early-time phenomena having negative values of time, and events post-peak having positive values of time. During Phase 1c, we observe the continued rise in modeled radius and decrease in modeled temperature. During this period, the increase of the fireball radius is now no longer fast enough to offset the cooling photosphere, and the result is a decreasing luminosity. The end of phase 1c occurs when the modeled supernova radius reaches its peak value.  \par 

The onset of Phase 2 is the time of the maximum modeled radius. Throughout Phase 1, the supernovae show growth in modeled radius, in keeping with an expanding, optically thick fireball. The turnover in the modeled fireball radius is consistent with the transition of the fireball photosphere from optically thick to optically thin. Our simple model of the temperature and radius of the fireball is valid in the optically thick regime, and the well-behaved patterns of temperature and radius in Phase 1 give support to its applicability. We expected to see evidence of the breakdown of the model at late times as the photosphere became transparent. Our hypothesis was that the transition to an optically thin supernova would coincide with emission that is weighted towards smaller radii and perhaps towards higher temperatures. This is consistent with our analysis of the Open Supernova Survey supernovae, exemplified in Figure \ref{4SQ}. The onset of Phase 2, the maximum modeled radius, is denoted by the rightmost vertical dashed line and is fairly contemporaneous with a change in progression of modeled temperature, here seen as a flattening or slight increase in temperature. Thus, the onset of Phase 2 marks a transition from a well-motivated optically-thick model of the physical properties (radius and temperature) of the fireball to a period where more complex analysis is required. These late-time temperatures and radii are not to be interpreted as a model of a physical fireball. For example, the decreasing modeled radii commonly observed in Phase 2, should not be interpreted as a shrinking photosphere, rather it is the result of applying the optically thick blackbody fireball model to a system requiring complex radiative transfer modeling beyond the scope of this paper. 

\subsection{Classification Types}

We group our analysis of supernovae according to the classification types from the Open Supernova Catalog. In the following sections we use the insight derived from our model to briefly describe the progression of these supernovae. The supernova type, count of supernovae, timings and physical parameters of onset, maximum temperature, luminosity, radius etc. are summarized in Table \ref{finsum}. The table displays these statistics by type presenting the median value, and showing offsets to the first (superscript) and third (subscript) quartile values. Following the discussion of each supernova classification type we then compare across types.

\begin{figure}[t]

\includegraphics[width=\linewidth]{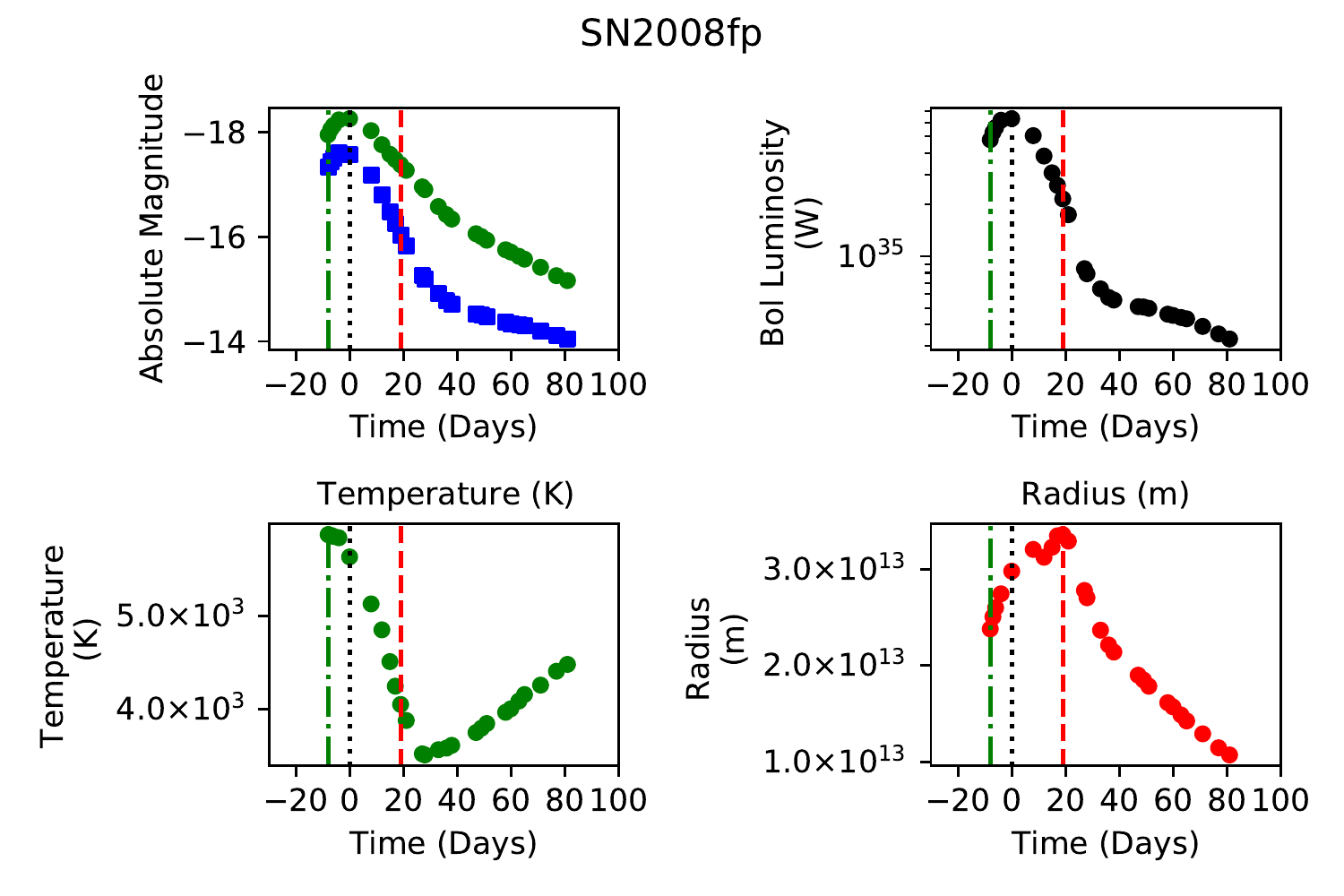}
\caption{SN2008fp is a Type II supernovae from our sample. Reference the caption of Figure \ref{4SQ} for more information regarding these plots.}
\label{Asymptotic Growth Type II}
\end{figure}
\subsubsection{Type Ia}

Our sample of Type Ia supernovae consists of 86 unique events. We observe some similarities in peak parameters, timings of phases, and behavior of the fireball. The values of the maximum modeled parameters and timings of phases are summarized in Table \ref{finsum}. For an example of a Type Ia supernovae, reference Figure \ref{4SQ} and its caption.

We observe a maximum temperature of  $1.0_{-0.7}^{+0.2} \times 10^4$ K around $11_{-2}^{+3}$ days before peak luminosity. We note that the maximum modeled luminosity is  $7.5_{-2.2}^{+2.0} \times 10^{35}$ W.  We observe these supernovae reaching their maximum modeled radius of $3.4_{-0.4}^{+0.4} \times 10^{13} $ meters about $17_{-4}^{+3}$ days after peak luminosity. We note the fireball expansion velocity of $9.5_{-1.4}^{+1.6} \times 10^{6} $ meters per second with radiative energies of around $2.7_{-0.6}^{+0.8} \times 10^{42}$ joules. The temperature at the maximum modeled radius is $4.4_{-0.2}^{+0.2}\times10^3 $K. This is approximately as expected since we hypothesized there would be a change in the opacity due to electron recombination as the fireball cooled. \par

\subsubsection{Type II}

Our sample of Type II supernovae contains 19 supernovae. This sample is smaller than our sample of Type Ia supernovae, but we can still observe uniformity in the maximum modeled parameters and timings of phases. These values are summarized in Table \ref{finsum}.

Similar to our Type Ia supernovae, we see most of the maximum modeled parameters and timings are similar across the sample of Type II supernovae. We observe maximum modeled temperatures of $9.4_{-1.6}^{+0.8} \times 10^{3}$ K about $10_{-4}^{+0}$ days before the maximum modeled luminosity. We observe a maximum modeled luminosity of  $1.8_{-1.2}^{+2.0} \times 10^{35} $ W. About $33_{-14}^{+13}$ days later, our Type II supernovae reach a maximum radius of  $2.0_{-0.5}^{+0.6} \times 10^{13}$ meters at a rate of $3.1_{-1.1}^{+1.5} \times 10^{6} $ meters per second. The total radiative energy is  $7.1_{-4.2}^{+4.9} \times 10^{41}$ joules. We observe a temperature at maximum radius, $4.1_{-0.1}^{+0.4} \times 10^3 $ K, that is consistent with opacity loss due to electron recombination.

\begin{figure}[t]

\includegraphics[width=\linewidth]{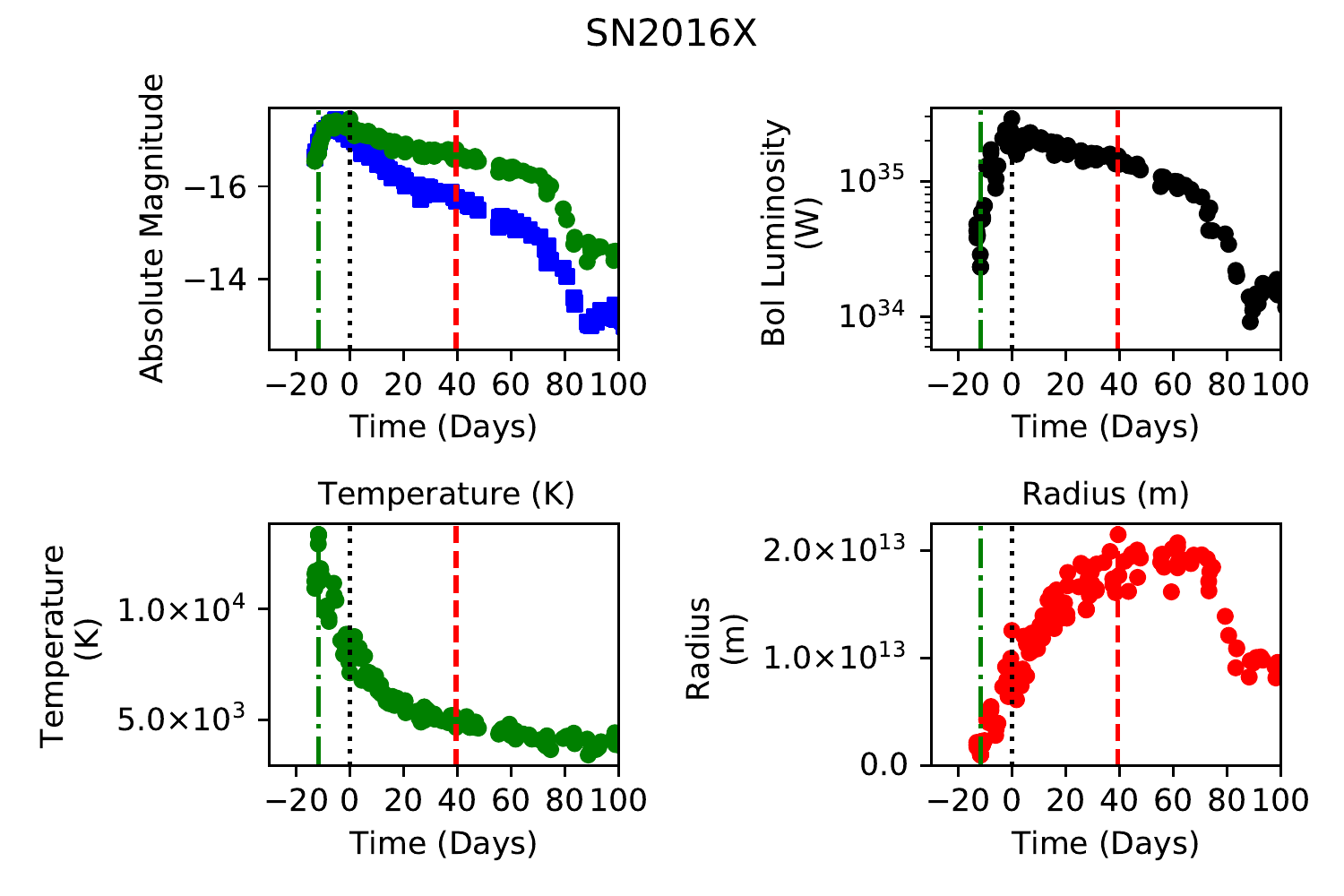}
\caption{SN2016X is a Type II P supernova from our sample. Reference the caption of Figure \ref{4SQ} for more information regarding these plots.}
\label{fig:AGIIP}
\end{figure}

\subsubsection{Type II P}

The 19 Type II P supernovae in our sample reach a maximum modeled temperature of  $1.0_{-0.2}^{+0.2} \times 10^4$ K about  $8_{-5}^{+5}$ before they reach their maximum modeled luminosity of $1.8_{-1.2}^{+2.0} \times 10^{35}$ W. These supernovae grew to a maximum radius of $1.7_{-0.6}^{+0.6} \times 10^{13}$ meters about $42_{-8}^{+11}$ days after the maximum modeled luminosity. The rate of growth for our Type II P supernovae is $2.3_{-0.8}^{+1.6} \times 10^{6}$ meters per second. We observe a total radiative energy of $5.2_{-2.7}^{+4.8} \times 10^{41}$ joules and a temperature at maximum radius, $4.2_{-0.1}^{+0.1} \times 10^3 $ K, that is consistent with recombination. 

\begin{figure}[t]

\includegraphics[width=\linewidth]{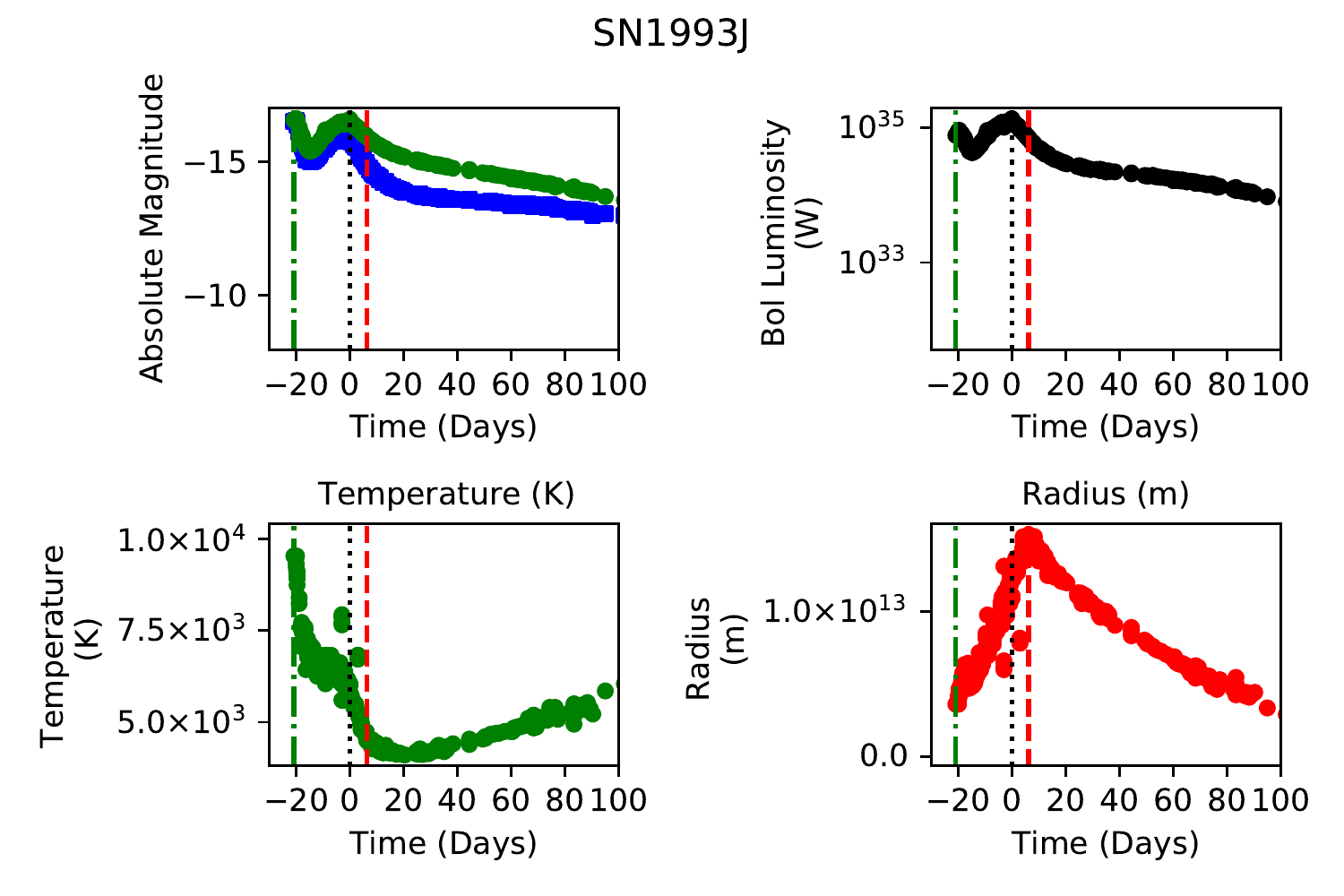}
\caption{SN1993J is an example of a Type IIb supernovae that exhibits the canonical double peaked behaviour in its luminosity light curves. In our sample of Type IIb supernovae, there are four that exhibit double peaks. Three of these, including SN1993J, show the maximum luminosity occurring during the second peak. It is unclear if this is due to a lack of early onset photometry or something intrinsic to these supernovae. Reference the caption of Figure \ref{4SQ} for more information regarding these plots. Compare to Figure \ref{fig:NCIIb} for an example of a Type IIb supernovae lacking this double peaked behavior.}
\label{fig:CIIb}
\end{figure}

\subsubsection{Type IIb}

\begin{figure}[t]

\includegraphics[width=\linewidth]{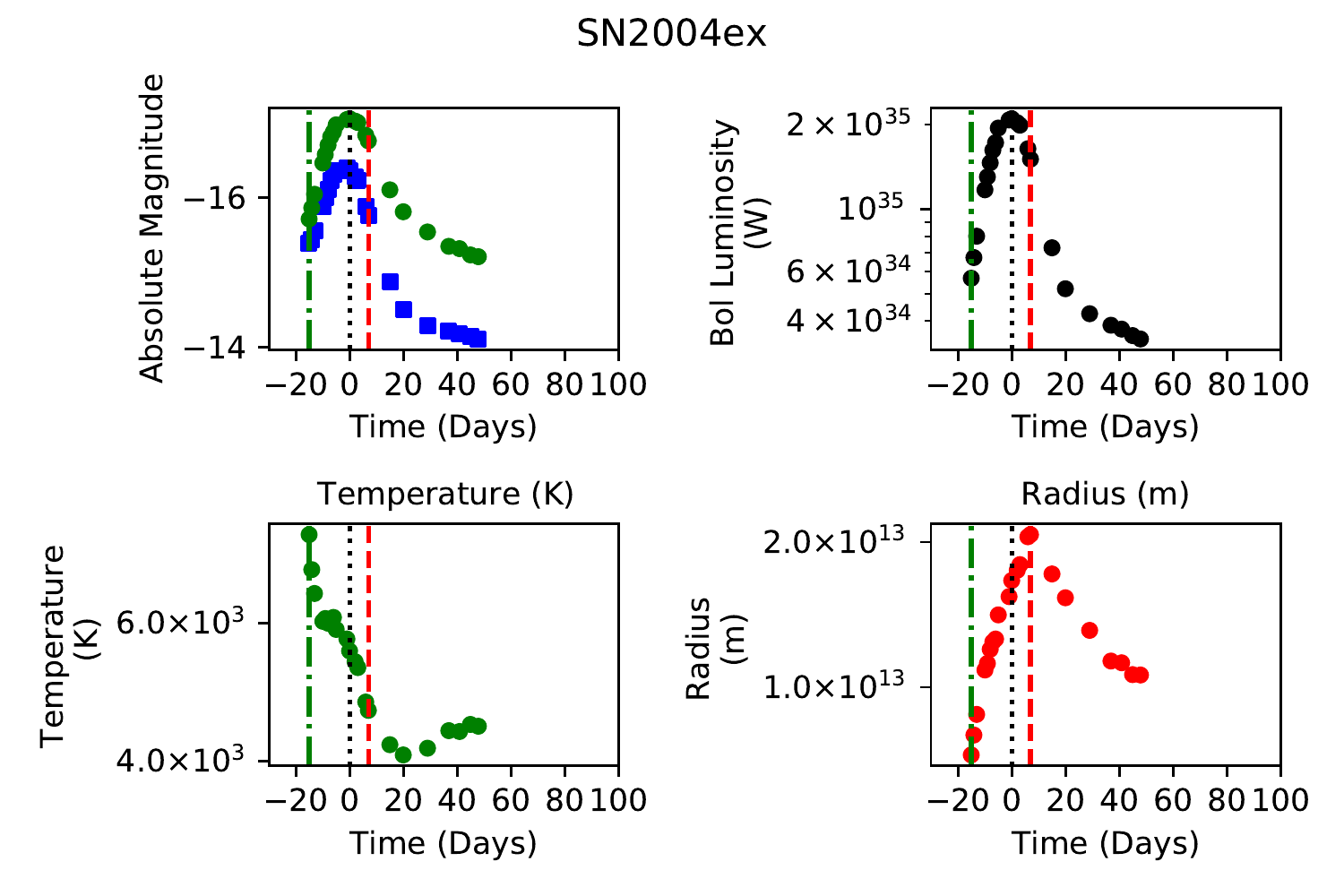}
\caption{SN2004ex is an example of Type IIb supernovae that does not exhibit the canonical double peaked behaviour in its luminosity light curves. Reference the caption of Figure \ref{4SQ} for more information regarding these plots. Compare to Figure \ref{fig:CIIb} for an example of a Type IIb supernovae showing the canonical double peaked behavior.}
\label{fig:NCIIb}
\end{figure}

We have a  sample of eleven Type IIb supernovae. Of the eleven, five exhibited the canonical behavior of two peaks in the luminosity light curve while six did not. Figure \ref{fig:CIIb} is an example of a canonical Type IIb supernova. The luminosity curves in both bands show evidence of double peaks although the supernova appears to have been detected shortly after the first peak. Figure \ref{fig:NCIIb} is a non canonical Type IIb, with luminosity curves that look similar to those of Type Ia, II, and II P supernovae.  \par

The Type IIb supernovae are less similar to each other than we see for other supernova classification types. We see differences in the modeled maximum parameters and their timings. We observe a modeled maximum temperature of $7.0_{-0.6}^{+1.4} \times 10^3 $ K around $10_{-6}^{+5}$ days before peak luminosity. We observe a modeled maximum luminosity of $1.4_{-6.7}^{+0.2} \times 10^{35}$ W. We observe a maximum radius of $1.7_{-0.5}^{+0.3} \times 10^{13}$ meters. These supernovae reach their modeled maximum radius about $7_{-2}^{+4}$  days after peak luminosity at a rate of $4.9_{-2.0}^{+2.8} \times 10^{6}$ meters per second. We approximate the total radiative energy to be $3.5_{-2.3}^{+1.1} \times 10^{41}$ joules. Similar to other classification types, the temperature at the modeled maximum radius is consistent with the temperature of recombination.

Despite our small sample of this supernova type, we notice some interesting differences in the behavior of the luminosity curves. Most compelling is the lack of the double peaked light curve in the non-canonical Type IIb supernovae. The canonical Type IIb supernovae reach their maximum temperature and their maximum modeled radius later than the non canonical sample. The cause of this discrepancy is not determined.

\subsection{Comparison}

We begin our comparison of these types by examining Table \ref{finsum}. The sample is heterogeneous with respect to supernovae count by type, with the plurality being of Type Ia (87) followed by Type II P (21), Type II (19), and Type IIb (11). The table follows the timeline of phases in our discussion and presents times and values for maximum modeled temperature, luminosity, radius, as well as temperature at maximum radius, the total radiative energy from the light-curve, and the fireball velocity as estimated from the modeled radial growth.

\begin{table}[h]
\resizebox{\textwidth}{!}{
\begin{tabular}{c c c c c c c c c c c c}
&&Phase 1a&&Phase 1b&Phase 1c($t \equiv 0$)&&Phase 2\\
Supernova Type&SN Count&Onset time&Max T&Time of max T&Max L(B band)&Max R&Time of max R&Temperature at max R&Radiative energy& Fireball Velocity\\
&&(Days)&(K)&(Days)&(W)&(m)&(Days)&(K)&(J)&(m/s)\\
\hline
&&&&&&&&&\\
Type Ia&86&$-19_{-2}^{+2}$& $1.0_{-0.7}^{+0.2} \times 10^4$ & $-11_{-2}^{+3}$ & $7.5_{-2.2}^{+2.0} \times 10^{35}$ & $3.4_{-0.4}^{+0.4} \times 10^{13}$ & $17_{-4}^{+3}$&$4.4_{-0.2}^{+0.2}\times10^3 $&$2.7_{-0.6}^{+0.8} \times 10^{42}$ & $9.5_{-1.4}^{+1.6} \times 10^{6} $  & \\
&&&&&&&&&\\

\hline
&&&&&&&&&\\
Type II&19&-14& $9.4_{-1.6}^{+0.8} \times 10^{3}$ & $-10_{-4}^{+0}$ & $1.8_{-1.2}^{+2.0} \times 10^{35}$ & $2.0_{-0.5}^{+0.6} \times 10^{13}$& $33_{-14}^{+13}$&$4.1_{-0.1}^{+0.4} \times 10^3$& $7.1_{-4.2}^{+4.9} \times 10^{41}$ & $3.1_{-1.1}^{+1.5} \times 10^{6} $ \\
&&&&&&&&&&\\

\hline
&&&&&&&&&&\\
Type II P&19&-13& $1.0_{-0.2}^{+0.2} \times 10^{4}$& $-8_{-5}^{+5}$& $1.1_{-6.4}^{+1.0} \times 10^{35}$ & $1.7_{-0.6}^{+0.6} \times 10^{13}$ & $42_{-8}^{+11}$ &$4.2_{-0.1}^{+0.1} \times10^3$& $5.2_{-2.7}^{+4.8} \times 10^{41}$ & $2.3_{-0.8}^{+1.6} \times 10^{6}$\\
&&&\\

\hline
&&&&&&&&&\\
Type IIb&11&-20& $7.0_{-0.6}^{+1.4} \times 10^{3}$ & $-10_{-6}^{+5}$ & $1.4_{-6.7}^{+0.2} \times 10^{35}$ & $1.7_{-0.5}^{+0.3} \times 10^{13}$ & $7_{-2}^{+4}$ &$4.5_{-0.2}^{+0.1}\times10^3$& $3.5_{-2.3}^{+1.1} \times 10^{41}$ & $4.9_{-2.0}^{+2.8} \times 10^{6}$\\
&&&&&&&&&&\\
\end{tabular}
}\caption{A summary of our meta-analysis of the Open Supernova Catalog containing modeled maximum values and timings of phases. The table is divided by supernova classification type. The values over a given supernovae type are given by quoting the median with the superscript denoting the offset to the third quartile and the subscript the offset to the first quartile. The first column gives the classification type. Next is the count of supernovae in each category. Those supernovae with a complete rise in luminosity have an estimated onset time. Next is the maximum modeled temperature and when it occurred. In the next column we give the maximum modeled luminosity. The time of this maximum is set to be 0 as is standard in the field. We then present the maximum modeled radius and its timing. We have included the temperature at maximum radius as this value is an indicator of a transition from an optically thick to an optically thin fireball. We note that this temperature of around 4500 K is consistent with free electrons recombining with ions (e.g. hydrogen), causing the fireball to lose an important source of opacity. We estimate the total radiative energy of the supernova, obtained by integrating the luminosity over time. The last column gives the fireball expansion velocity in meters per second. This value is approximated from the slope of the radial growth. Some differences across the types are the maximum modeled temperature, luminosity, and the time of maximum modeled radius. For example, we see the maximum modeled temperature, luminosity, and fireball expansion velocity are highest for Type Ia supernovae, as expected. }
\label{finsum}
\end{table}

Onset times are most clearly defined for Type Ia supernovae, with the estimated onset time being roughly 15 days before peak luminosity. Most of our Type II supernova were detected around the time of peak luminosity, leading to a lack of estimated onset times, though we identify a subset of Type II supernovae with onset times approximately 5 days before peak luminosity.\par 

The time of maximum temperature for Type Ia supernovae is about 5 days before peak luminosity, while these two events occur around the same time for Type II supernovae. The maximum temperature is roughly 10,000K for all types except for the Type IIb supernovae which are slightly lower. Our Type Ia temperature curves show these events heat up and cool quicker than Type II and II P supernovae. This is supported by the earlier time of maximum temperature, and an earlier time of maximum modeled radius. \par

There is a difference in approximated maximum luminosity, with Type Ia being several times more luminous than the other types (8.5x for the ratio between Type Ia and Type II P and Type IIb, 3.0x for Type II). Type Ia supernovae reach a maximum radius that is about twice as large as the other types. We also see that they grow faster, with an earlier time of maximum radius. The time of maximum radius for Type Ia supernovae is roughly 27 days after peak luminosity, while for Type II and Type II P supernovae this value is roughly 70 days.

Type IIb supernovae lack a hydrogen envelope and show some similarity with Type Ia supernovae in their progression (compare Figure \ref{4SQ} (Type Ia) with Figure \ref{fig:NCIIb} (Type IIb)). These types share earlier times of maximum temperature and radius, albeit with Type IIb supernovae having lower peak luminosity, radiative energy and maximum fireball radius. As we discussed earlier, it was expected that the fireball model would produce a maximum modeled radius as the fireball loses opacity and becomes optically thin. The lack of a hydrogen envelope in these supernovae may play a role in the earlier transition from an optically thick fireball to an optically thin one. \par

\section{Conclusion}

We developed a data pipeline that pulls photometric data from the Open Supernova Catalog and applies a simple optically thick spherically symmetric fireball model to examine the behavior of these events over time. The model led us to characterize distinct phases throughout their development. We labeled the continuous modeled growth of the fireball Phase I. This behaviour is consistent with our assumption of emission from an optically thick source, and sees: the increase in luminosity, temperature and radius in Phase 1a; the onset of a decrease in temperature in Phase 1b; and the onset of a decrease in luminosity in Phase 1c. We see the model grow to a maximum radius and begin to shrink in size. We denote this as the onset of Phase II. This is a turning point when our model is no longer describing an opaque outermost photosphere and instead we are likely seeing into the fireball to some depth. We note that this occurs uniformly at a modeled temperature of around 4,500 K, consistent with the loss of electron scattering opacity due to recombination. \par

Although they grow differently, the maximum modeled radius of all the supernovae classification types in our survey is roughly $10^{13}$ meters.  The peak modeled temperatures are similarly uniform at approximately $10^{4}$ K, with the exception of the handful of Type IIb supernovae whose maximum temperatures are a few thousand K cooler. There are differences in timing, with Type Ia supernovae reaching maximum temperature a few days before peak luminosity where this is roughly contemporaneous for Type II supernovae. We see a difference in the time of maximum modeled radius. Type Ia supernovae reach their maximum radius quicker than the other types, which is consistent with their higher fireball velocities. We also see Type Ia supernovae reach higher luminosities, approximately 8.5 times higher than the other types in the survey. \par

By applying our model to these events, we note supernovae have some common photosphere behaviors over their duration. There is a rise in all of these parameters until maximum temperature, followed by continual growth and cooling until maximum luminosity, then they continue to cool and dim until reaching a maximum modeled radius. The maximum modeled radius across all types is on average about the size of our solar system, with a surface temperature equivalent to the surface of the Sun. There is space for further work in applying simple models to the broad set of supernovae light curves available from catalogs like the Open Supernova Catalog. Future work could include constructing a more complex physically motivated model that accounts for differences in fireball geometry and includes simplified radiative transfer to go beyond the assumption of an optically thick photosphere. This work would also benefit from corrections for extinction mapping using the galactic coordinates of each supernovae to fully account for flux that is absorbed as photons travel through space.

\begin{acknowledgments}
This work was supported in part by the Sonoma State University Department of Physics \& Astronomy's Hichwa/McQuillen Summer Undergraduate Research Award. 
\end{acknowledgments}

\section{Appendix}

This appendix presents tables that contain summary values for each individual supernova's modeled parameters and the timing of their phases. The tables are separated by classification type and are sorted alphabetically. The values in these tables are used in the statistical summary provided in Table 1 in the main body of this paper. 

In these appendix tables, we give: the maximum values of the modeled temperature, luminosity, and radius of each supernova in our sample as well as their timing. The times presented represent the transition from one phase to another. Our phases begin at the onset of the supernova (Phase 1a), transitioning to the next phase at the modeled maximum values of temperature (Phase 1b), luminosity (Phase 1c) and radius (Phase 2). The phases are indicated in the header above the columns that give their corresponding
values. 

A second appendix containing plots of absolute magnitude, luminosity, temperature, and radius for each supernovae is available upon request from the author (\url{mailto:jacob.marshall.astr@gmail.com}).

The Open Supernova Catalog often contains photometry from multiple sources for a given supernova. We have included the photometric sources (in the B and V bands) for each supernova as tables by supernova classification type. We present these references in the format provided by the Open Supernova Catalog:
ADS bibliographic codes (bibcodes). See the Astrophysics Data System (\url{https://ui.adsabs.harvard.edu}) for details. 

\begin{table}[t]
    \centering 
    \includegraphics[width=\textwidth]{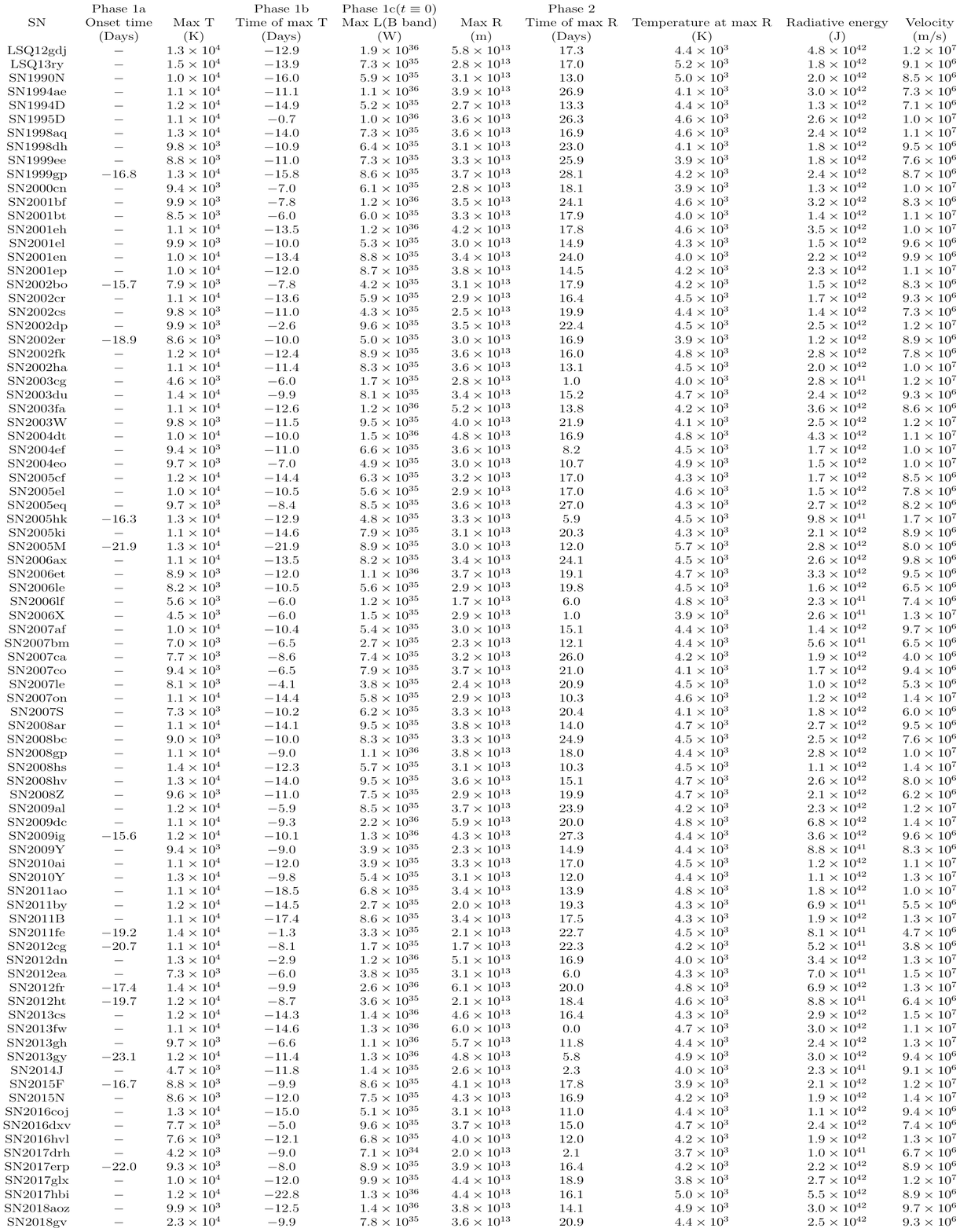}
    \caption{This table contains summary values for all of our Type Ia supernovae. The supernova name is followed by its estimated onset time, if one is determined as discussed in the main body of the paper. The next two columns are the maximum modeled temperature (Kelvin) and its timing (days). Next is the maximum modeled B band luminosity (Watts). This time of maximum luminosity is defined as 0 as standard in the field. The next two columns give the maximum modeled radius (meters) and its timing. We have included the temperature at maximum radius as this value is an indicator of a transition from an optically thick to an optically thin fireball. We note that this temperature of around 4500 degrees kelvin is consistent with free electrons recombining with ions (e.g. hydrogen), causing the fireball to lose an important source of opacity.  Following that is an estimate of the total radiative energy (Joules) and the fireball expansion velocity (meters per second). The values presented in this table are summarized by supernova type in the body of the main paper.}
    \label{tab:LGIa}
\end{table}

\begin{table}[t]
\tiny
\resizebox{\textwidth}{\height}{
\begin{tabular}{c  p{\paperwidth}}
SN&Sources\\
iPTF13dge&2014Ap\&SS.354...89B, 2016A\&A...592..A40F\\
LSQ12gdj&2015ApJS..219...13W, 2015A\&A...579A..40S, 2014Ap\&SS.354...89B\\
LSQ13ry&2015ApJS..219...13W, 2014Ap\&SS.354...89B\\
SN1990N&1998AJ....115..234L,SUSPECT, 1998AJ....115..234L,Sternberg Astronomical Institute Supernova Light Curve Catalogue\\
SN1994ae&2005ApJ...627..579R,CfA Supernova Archive, 1999AJ....117..707R,CfA Supernova Archive, 1994IAUC.6115....1V,Sternberg Astronomical Institute Supernova Light Curve Catalogue, 1989IAUC.6115....1P,Sternberg Astronomical Institute Supernova Light Curve Catalogue, 2001PASP..113.1349H,Sternberg Astronomical Institute Supernova Light Curve Catalogue, 1997PAZh...23...30T,Sternberg Astronomical Institute Supernova Light Curve Catalogue, 2004MNRAS.349.1344A\\
SN1994D&1995AJ....109.2121R,Sternberg Astronomical Institute Supernova Light Curve Catalogue, La Palma, supernova archieve,Sternberg Astronomical Institute Supernova Light Curve Catalogue, 1994IAUC.5951....1C,Sternberg Astronomical Institute Supernova Light Curve Catalogue, 1994IAUC.5950....1W,Sternberg Astronomical Institute Supernova Light Curve Catalogue, 1994IAUC.5952....3H,Sternberg Astronomical Institute Supernova Light Curve Catalogue, F.Patat et al, ESO Preprint, No.1100, 1995,Sternberg Astronomical Institute Supernova Light Curve Catalogue, 1994IAUC.5976....3A,Sternberg Astronomical Institute Supernova Light Curve Catalogue, 1995PAZh...21..678T,Sternberg Astronomical Institute Supernova Light Curve Catalogue, 1994IAUC.5958....3M,Sternberg Astronomical Institute Supernova Light Curve Catalogue, 2004MNRAS.349.1344A\\
SN1995D&1989IAUC.6166....1M,Sternberg Astronomical Institute Supernova Light Curve Catalogue, 1999AJ....117..707R,CfA Supernova Archive, 2004MNRAS.349.1344A, 2001PASP..113.1349H,Sternberg Astronomical Institute Supernova Light Curve Catalogue\\
SN1998aq&2005ApJ...627..579R,CfA Supernova Archive\\
SN1998dh&2012MNRAS.425.1789S,2010ApJS..190..418G, 2006AJ....131..527J,CfA Supernova Archive\\
SN1999ee&2002AJ....124.2100S,SUSPECT, 2002AJ....124.2100S,Sternberg Astronomical Institute Supernova Light Curve Catalogue\\
SN1999gp&2010ApJS..190..418G,2012MNRAS.425.1789S, 2001AJ....122.1616K, 2006AJ....131..527J,CfA Supernova Archive\\
SN2000cn&2012MNRAS.425.1789S,2010ApJS..190..418G, 2006AJ....131..527J,CfA Supernova Archive\\
SN2001bf&2012MNRAS.425.1789S,2010ApJS..190..418G, 2009ApJ...700..331H,Sternberg Astronomical Institute Supernova Light Curve Catalogue, VSNET 2001,Sternberg Astronomical Institute Supernova Light Curve Catalogue\\
SN2001bt&2004AJ....128.3034K\\
SN2001eh&2012MNRAS.425.1789S,2010ApJS..190..418G, 2009ApJ...700..331H,Sternberg Astronomical Institute Supernova Light Curve Catalogue\\
SN2001el&2003AJ....125..166K,Sternberg Astronomical Institute Supernova Light Curve Catalogue, 2003AJ....125..166K\\
SN2001en&2012MNRAS.425.1789S,2010ApJS..190..418G, 2009ApJ...700..331H,Sternberg Astronomical Institute Supernova Light Curve Catalogue\\
SN2001ep&2012MNRAS.425.1789S,2010ApJS..190..418G, 2009ApJ...700..331H,Sternberg Astronomical Institute Supernova Light Curve Catalogue\\
SN2001V&2009ApJ...700..331H,Sternberg Astronomical Institute Supernova Light Curve Catalogue, VSNET 2001,Sternberg Astronomical Institute Supernova Light Curve Catalogue, 2003A\&A...397..115V,SUSPECT, 2003A\&A...397..115V,Sternberg Astronomical Institute Supernova Light Curve Catalogue, 2006AJ....132.2024L, 2012MNRAS.425.1789S,2010ApJS..190..418G\\
SN2002bo&2012MNRAS.425.1789S,2010ApJS..190..418G, 2004MNRAS.348..261B,SUSPECT, 2004MNRAS.348..261B,Sternberg Astronomical Institute Supernova Light Curve Catalogue, 2009ApJ...700..331H,Sternberg Astronomical Institute Supernova Light Curve Catalogue, 2003A\&A...408..915S,Sternberg Astronomical Institute Supernova Light Curve Catalogue, 2004AJ....128.3034K, VSNET Oct2003,Sternberg Astronomical Institute Supernova Light Curve Catalogue\\
SN2002cr&2012MNRAS.425.1789S,2010ApJS..190..418G, VSNET Oct2003,Sternberg Astronomical Institute Supernova Light Curve Catalogue, 2009ApJ...700..331H,Sternberg Astronomical Institute Supernova Light Curve Catalogue\\
SN2002cs&2012MNRAS.425.1789S,2010ApJS..190..418G, VSNET Oct2003,Sternberg Astronomical Institute Supernova Light Curve Catalogue\\
SN2002dp&2012MNRAS.425.1789S,2010ApJS..190..418G, 2009ApJ...700..331H,Sternberg Astronomical Institute Supernova Light Curve Catalogue\\
SN2002er&2012MNRAS.425.1789S,2010ApJS..190..418G, 2004MNRAS.355..178P,Sternberg Astronomical Institute Supernova Light Curve Catalogue, 2003A\&A...401..479C,Sternberg Astronomical Institute Supernova Light Curve Catalogue\\
SN2002fk&2012MNRAS.425.1789S,2010ApJS..190..418G, VSNET Oct2003,Sternberg Astronomical Institute Supernova Light Curve Catalogue, 2009ApJ...700..331H,Sternberg Astronomical Institute Supernova Light Curve Catalogue\\
SN2002ha&2012MNRAS.425.1789S,2010ApJS..190..418G, VSNET Oct2003,Sternberg Astronomical Institute Supernova Light Curve Catalogue, 2009ApJ...700..331H,Sternberg Astronomical Institute Supernova Light Curve Catalogue\\
SN2003du&2005A\&A...429..667A,Sternberg Astronomical Institute Supernova Light Curve Catalogue, 2007A\&A...469..645S, 2009ApJ...700..331H,Sternberg Astronomical Institute Supernova Light Curve Catalogue, 2012MNRAS.425.1789S,2010ApJS..190..418G\\
SN2003fa&2012MNRAS.425.1789S,2010ApJS..190..418G, 2009ApJ...700..331H,Sternberg Astronomical Institute Supernova Light Curve Catalogue\\
SN2003W&2012MNRAS.425.1789S,2010ApJS..190..418G, 2009ApJ...700..331H,Sternberg Astronomical Institute Supernova Light Curve Catalogue\\
SN2004dt&2012MNRAS.425.1789S,2010ApJS..190..418G, 2010AJ....139..519C\\
SN2004ef&2010AJ....139..519C, 2012MNRAS.425.1789S,2010ApJS..190..418G, 2009ApJ...700..331H,Sternberg Astronomical Institute Supernova Light Curve Catalogue\\
SN2004eo&2010AJ....139..519C, 2007MNRAS.377.1531P,Sternberg Astronomical Institute Supernova Light Curve Catalogue, 2012MNRAS.425.1789S,2010ApJS..190..418G\\
SN2005cf&2009ApJ...700..331H,Sternberg Astronomical Institute Supernova Light Curve Catalogue, 2012MNRAS.425.1789S,2010ApJS..190..418G, 2007MNRAS.376.1301P, 2014Ap\&SS.354...89B\\
SN2005el&2010AJ....139..519C, 2012MNRAS.425.1789S,2010ApJS..190..418G, 2009ApJ...700..331H,Sternberg Astronomical Institute Supernova Light Curve Catalogue\\
SN2005eq&2012MNRAS.425.1789S,2010ApJS..190..418G, 2009ApJ...700..331H,Sternberg Astronomical Institute Supernova Light Curve Catalogue, 2010AJ....139..519C\\
SN2005hk&2019MNRAS.490.3882S,2012MNRAS.425.1789S, 2009ApJ...700..331H,Sternberg Astronomical Institute Supernova Light Curve Catalogue, 2014Ap\&SS.354...89B, 2008ApJ...680..580S,Sternberg Astronomical Institute Supernova Light Curve Catalogue\\
SN2005ki&2009ApJ...700..331H,Sternberg Astronomical Institute Supernova Light Curve Catalogue, 2019MNRAS.490.3882S,2012MNRAS.425.1789S, 2010AJ....139..519C\\
SN2005M&2012MNRAS.425.1789S,2010ApJS..190..418G, 2010AJ....139..519C, 2009ApJ...700..331H,Sternberg Astronomical Institute Supernova Light Curve Catalogue\\
SN2006ax&2010AJ....139..519C, 2009ApJ...700..331H,Sternberg Astronomical Institute Supernova Light Curve Catalogue\\
SN2006et&2010AJ....139..519C, 2009ApJ...700..331H,Sternberg Astronomical Institute Supernova Light Curve Catalogue\\
SN2006le&2012MNRAS.425.1789S,2010ApJS..190..418G, 2009ApJ...700..331H,Sternberg Astronomical Institute Supernova Light Curve Catalogue\\
SN2006lf&2012MNRAS.425.1789S,2010ApJS..190..418G, 2009ApJ...700..331H,Sternberg Astronomical Institute Supernova Light Curve Catalogue\\
SN2006X&2014Ap\&SS.354...89B, 2012MNRAS.425.1789S,2010ApJS..190..418G, 2010AJ....139..519C, 2009ApJ...700..331H,Sternberg Astronomical Institute Supernova Light Curve Catalogue\\
SN2007af&2014Ap\&SS.354...89B, 2009ApJ...700..331H,Sternberg Astronomical Institute Supernova Light Curve Catalogue, 2012MNRAS.425.1789S,2010ApJS..190..418G, 2010AJ....139..519C\\
SN2007bm&2009ApJ...700..331H,Sternberg Astronomical Institute Supernova Light Curve Catalogue, 2019MNRAS.490.3882S,2012MNRAS.425.1789S, 2014Ap\&SS.354...89B, 2010AJ....139..519C\\
SN2007ca&2009ApJ...700..331H,Sternberg Astronomical Institute Supernova Light Curve Catalogue, 2010AJ....139..519C, 2012MNRAS.425.1789S,2010ApJS..190..418G\\
SN2007co&2012MNRAS.425.1789S,2010ApJS..190..418G, 2009ApJ...700..331H,Sternberg Astronomical Institute Supernova Light Curve Catalogue, 2014Ap\&SS.354...89B\\
SN2007le&2010AJ....139..519C, 2012MNRAS.425.1789S,2010ApJS..190..418G, 2012ApJS..200...12H\\
SN2007on&2014Ap\&SS.354...89B, 2010AJ....139..519C\\
SN2007S&2010AJ....139..519C, 2014Ap\&SS.354...89B, 2009ApJ...700..331H,Sternberg Astronomical Institute Supernova Light Curve Catalogue\\
SN2008ar&2012MNRAS.425.1789S,2010ApJS..190..418G, 2012ApJS..200...12H\\
SN2008bc&2010AJ....139..519C\\
SN2008gp&2010AJ....139..519C, 2019MNRAS.490.3882S,2012MNRAS.425.1789S\\
SN2008hs&2019MNRAS.490.3882S,2012MNRAS.425.1789S, 2012ApJ...749...18B, 2014Ap\&SS.354...89B, 2012ApJS..200...12H\\
SN2008hv&2012ApJS..200...12H, 2014Ap\&SS.354...89B, 2012ApJ...749...18B, 2010AJ....139..519C\\
SN2008Z&2012ApJS..200...12H, 2012MNRAS.425.1789S,2010ApJS..190..418G\\
SN2009al&2012ApJS..200...12H, 2019MNRAS.490.3882S,2012MNRAS.425.1789S\\
SN2009dc&2012ApJS..200...12H, http://kanatatmp.g.hatena.ne.jp/kanataobslog/,Sternberg Astronomical Institute Supernova Light Curve Catalogue, 2011MNRAS.410..585S,Sternberg Astronomical Institute Supernova Light Curve Catalogue, 2019MNRAS.490.3882S,2012MNRAS.425.1789S, 2010AJ....139..519C, 2014Ap\&SS.354...89B\\
SN2009ig&2019MNRAS.490.3882S,2012MNRAS.425.1789S, 2014Ap\&SS.354...89B, 2012ApJ...749...18B, 2012ApJS..200...12H\\
SN2009Y&2014Ap\&SS.354...89B, 2012ApJS..200...12H\\
SN2010ai&2012ApJS..200...12H\\
SN2010Y&2012ApJ...749...18B, 2014Ap\&SS.354...89B, 2012ApJS..200...12H\\
SN2011ao&2014Ap\&SS.354...89B\\
SN2011by&2019MNRAS.490.3882S,2012MNRAS.425.1789S, 2014Ap\&SS.354...89B\\
SN2011B&2014Ap\&SS.354...89B\\
SN2011fe&2014Ap\&SS.354...89B, 2019MNRAS.490.3882S,2012MNRAS.425.1789S, 2013CoSka..43...94T,Sternberg Astronomical Institute Supernova Light Curve Catalogue, 2013NewA...20...30M, 2012JAVSO..40..872R,Sternberg Astronomical Institute Supernova Light Curve Catalogue\\
SN2012cg&2019MNRAS.490.3882S,2012MNRAS.425.1789S, 2014Ap\&SS.354...89B, 2018PASP..130f4101V, 2013NewA...20...30M\\
SN2012dn&2014MNRAS.443.1663C, 2014Ap\&SS.354...89B, 2016PASJ...68...68Y, 2019MNRAS.490.3882S,2012MNRAS.425.1789S\\
SN2012ea&2019MNRAS.490.3882S,2012MNRAS.425.1789S\\
SN2012fr&2014Ap\&SS.354...89B, 2017MNRAS.472.3437G, 2014AJ....148....1Z, 2018arXiv180310095C\\
SN2012ht&2018PASP..130f4101V, 2018arXiv180906381B, 2014Ap\&SS.354...89B\\
SN2013cs&2017MNRAS.472.3437G, 2015ApJS..219...13W, 2014Ap\&SS.354...89B\\
SN2013fw&2019MNRAS.490.3882S,2012MNRAS.425.1789S, 2014Ap\&SS.354...89B\\
SN2013gh&2016A\&A...592..A40F, 2019MNRAS.490.3882S,2012MNRAS.425.1789S, 2014Ap\&SS.354...89B\\
SN2013gy&2019MNRAS.490.3882S,2012MNRAS.425.1789S, 2017MNRAS.472.3437G, 2014Ap\&SS.354...89B\\
SN2014J&2014MNRAS.443.2887F,Sternberg Astronomical Institute Supernova Light Curve Catalogue, 2014ApJ...784L..12G,Sternberg Astronomical Institute Supernova Light Curve Catalogue, 2019MNRAS.490.3882S,2012MNRAS.425.1789S, 2014AJ....148....1Z, 2014CoSka..44...67T,Sternberg Astronomical Institute Supernova Light Curve Catalogue, 2018PASP..130f4101V\\
SN2015F&2017ApJ...850..111N, 2018arXiv180906381B, 2017MNRAS.472.3437G, 2014Ap\&SS.354...89B\\
SN2015N&2019MNRAS.490.3882S,2012MNRAS.425.1789S\\
SN2016coj&2019MNRAS.490.3882S,2012MNRAS.425.1789S\\
SN2016dxv&CPCS Alert 26244,Cambridge Photometric Calibration Server\\
SN2016hvl&2019MNRAS.490.3882S,2012MNRAS.425.1789S, CPCS Alert 26431,Cambridge Photometric Calibration Server\\
SN2017drh&2019MNRAS.490.3882S,2012MNRAS.425.1789S, CPCS Alert 26694,Cambridge Photometric Calibration Server\\
SN2017erp&2014Ap\&SS.354...89B, 2019MNRAS.490.3882S,2012MNRAS.425.1789S, CPCS Alert 28505,Cambridge Photometric Calibration Server\\
SN2017glx&2019MNRAS.490.3882S, 2012MNRAS.425.1789S\\
SN2017hbi&2019MNRAS.490.3882S,2012MNRAS.425.1789S\\
SN2018aoz&2014Ap\&SS.354...89B, 2019MNRAS.490.3882S,2012MNRAS.425.1789S\\
SN2018gv&2019MNRAS.490.3882S,2012MNRAS.425.1789S\\

\end{tabular}
}
\caption{List of photometric sources for each Type Ia supernovae as provided by the Open Supernova Catalog formatted as ADS bibcodes.}
\label{tab:LGIaSource}
\end{table}

\begin{table}[h]
    \centering
    \includegraphics[width=\textwidth]{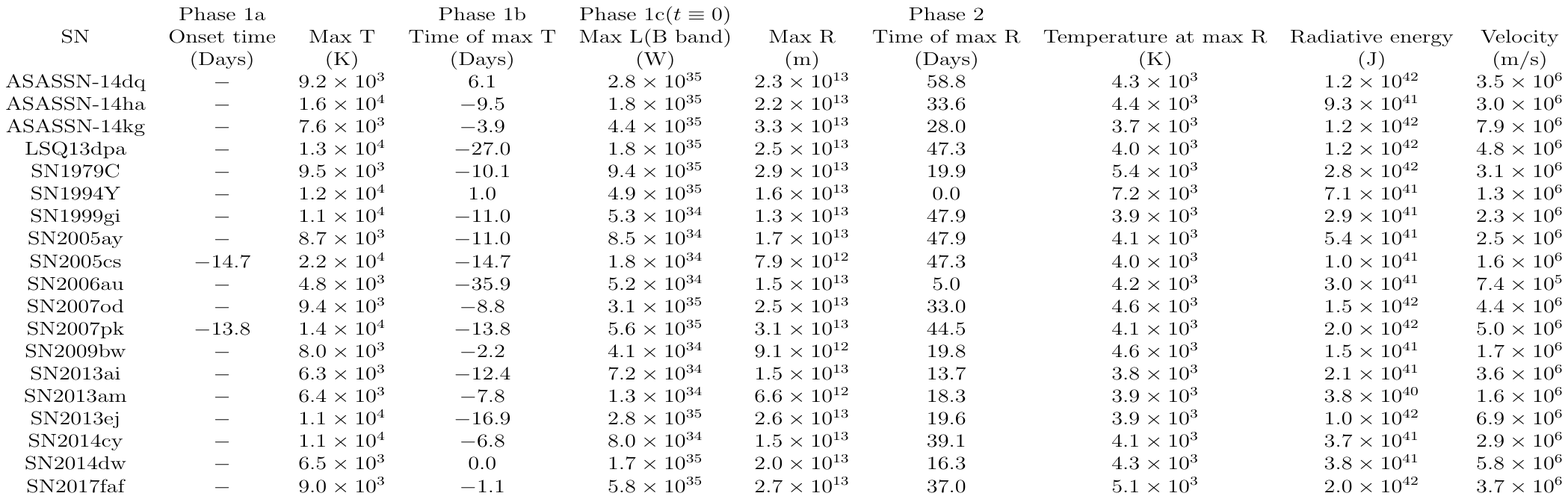}
    \caption{This table contains summary values for our Type II supernovae. Reference the caption of Table \ref{tab:LGIa} for more information on the contents of this table.}
    \label{tab:EGII}
\end{table}

\begin{table}[h]
    \centering
    \includegraphics[width=\textwidth]{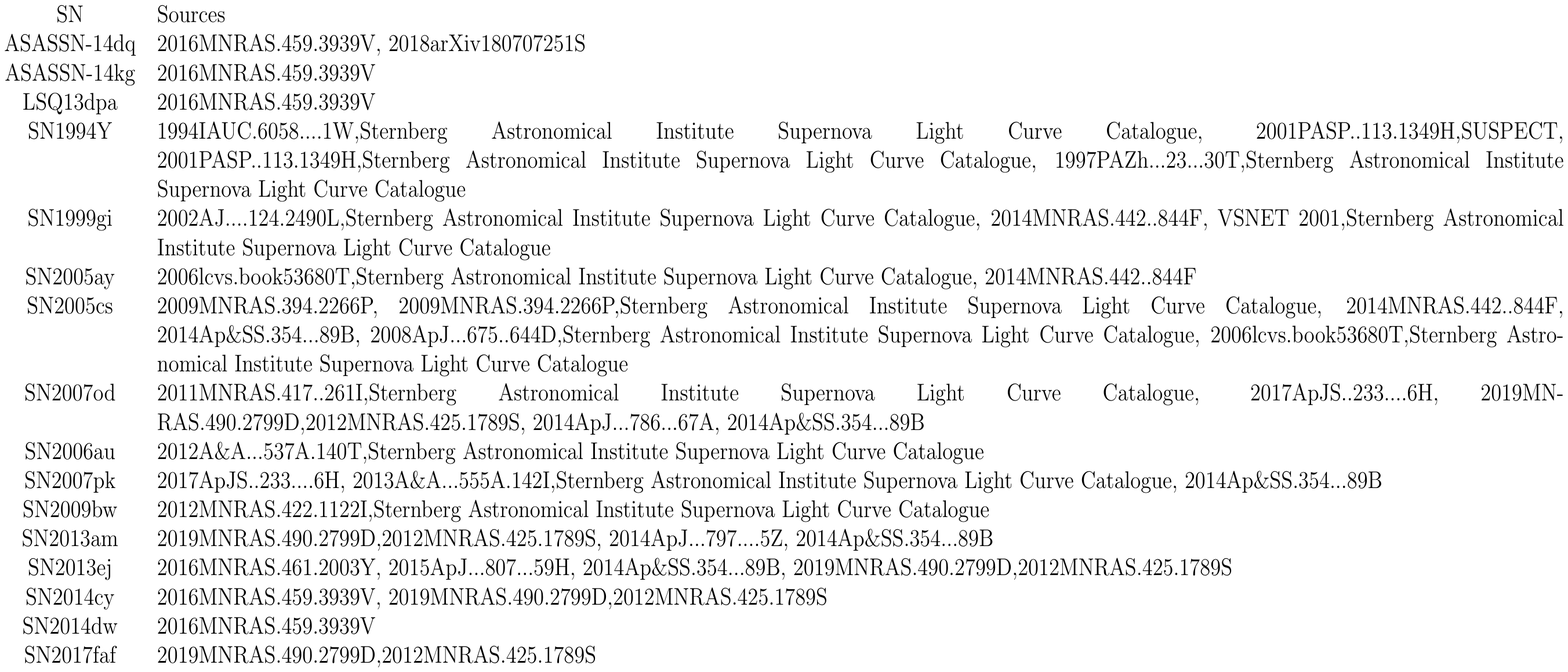}
    \caption{List of photometric sources for each Type II supernovae as provided by the Open Supernova Catalog formatted as ADS bibcodes.}
    \label{tab:EGIIsource}
\end{table}

\begin{table}[h]
    \centering
    \includegraphics[width=\textwidth]{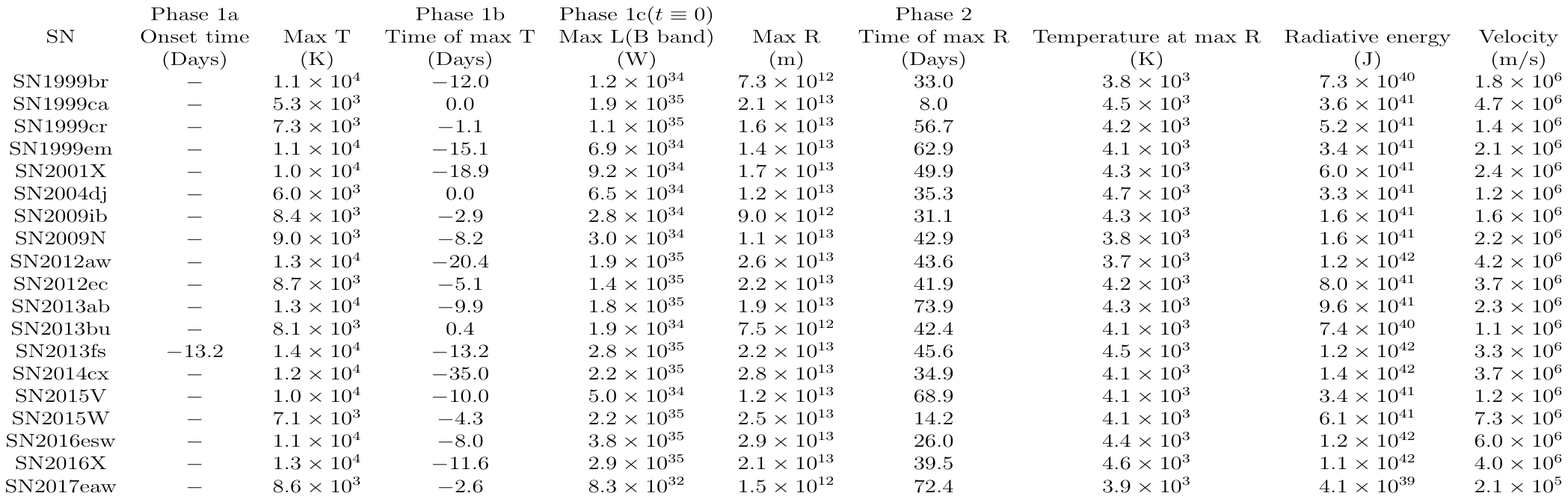}
    \caption{This table contains summary values for our Type II P supernovae. Reference the caption of Table \ref{tab:LGIa} for more information on the contents of this table.}
    \label{tab:AIIP}
\end{table}

\begin{table}[h]
    \centering
    \includegraphics[width=\textwidth]{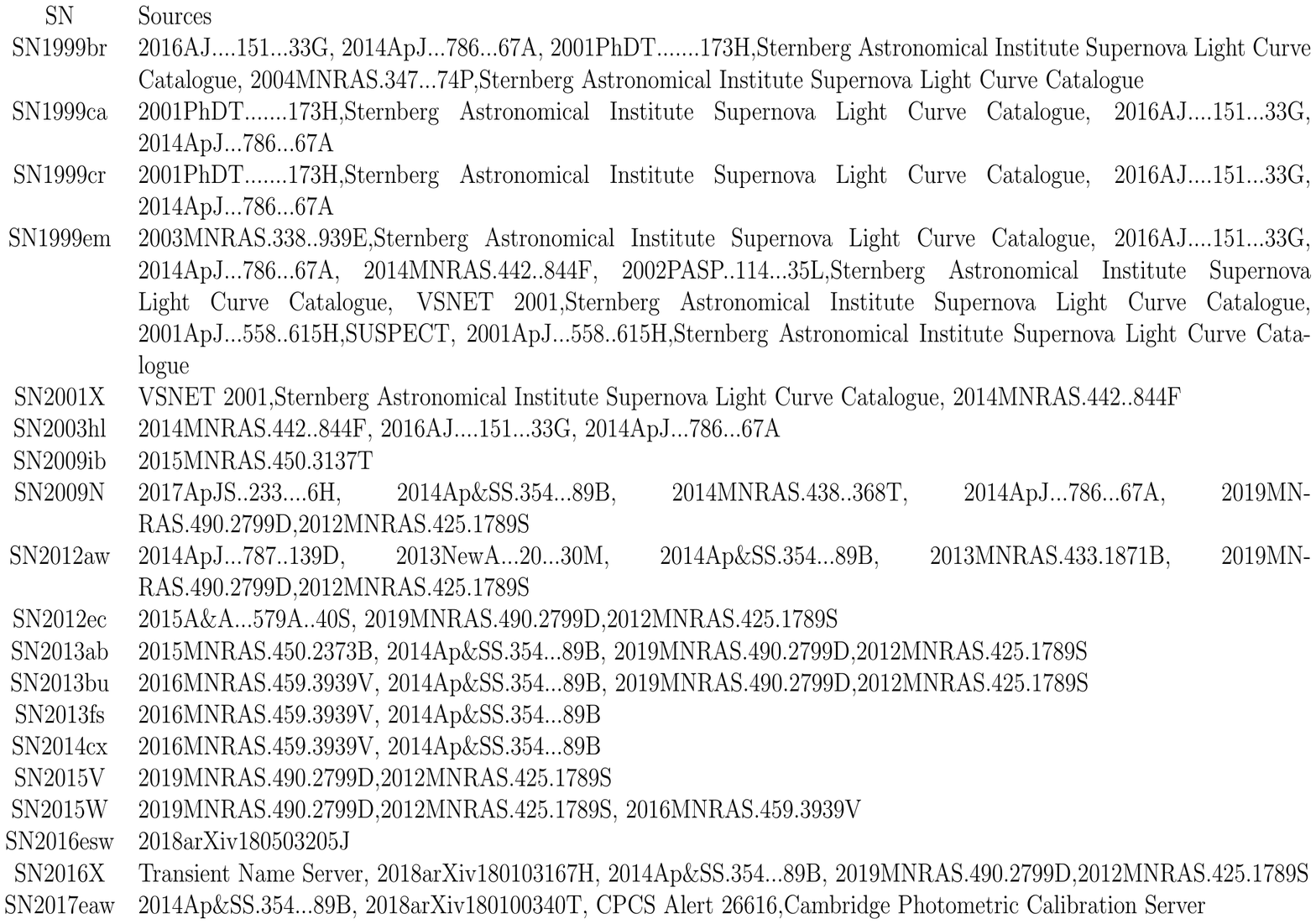}
    \caption{List of photometric sources for each Type II P supernovae as provided by the Open Supernova Catalog formatted as ADS bibcodes.}
    \label{tab:AIIPSource}
\end{table}

\begin{table}[h]
    \centering
    \includegraphics[width=\textwidth]{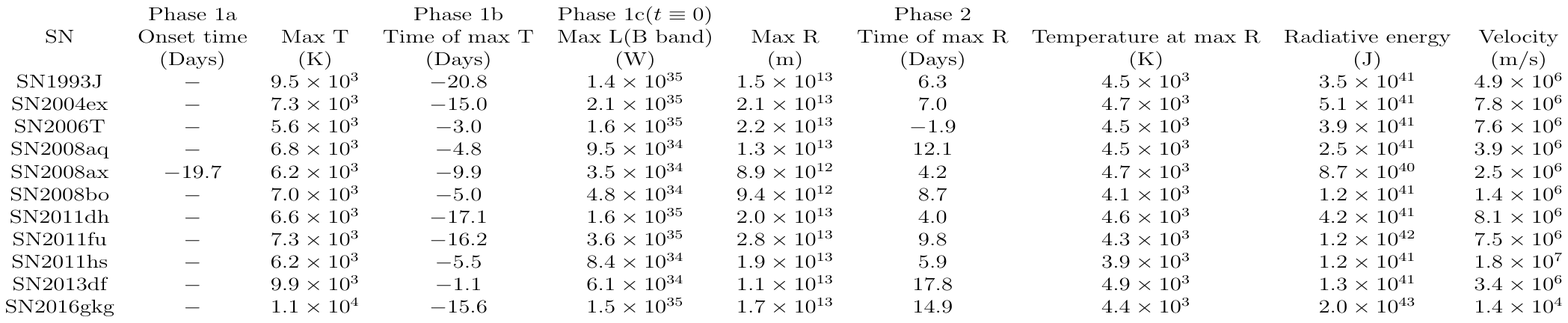}
    \caption{This table contains summary values for our Type IIb supernovae. Reference the caption of Table \ref{tab:LGIa} for more information on the contents of this table.}
    \label{tab:IIb}
\end{table}

\begin{table}[t]
    \centering
    \includegraphics[width=\textwidth]{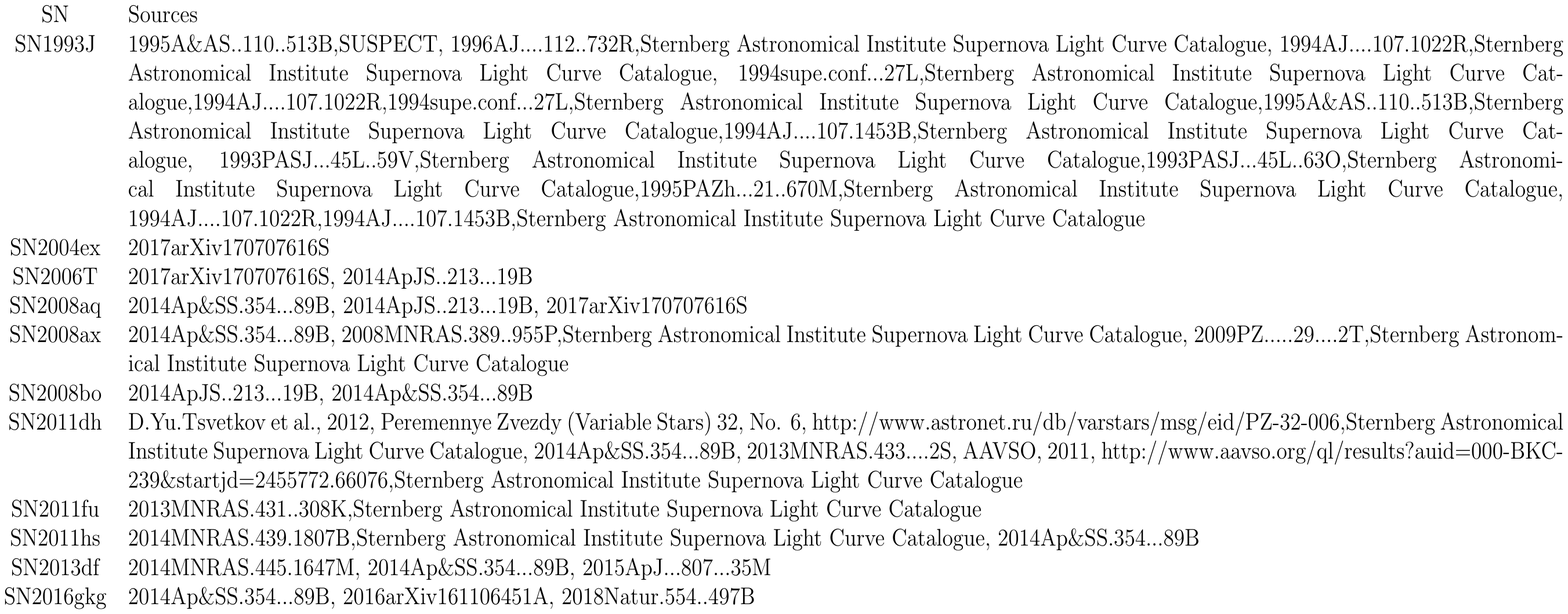}
    \caption{List of photometric sources for each Type IIb supernovae as provided by the Open Supernova Catalog formatted as ADS bibcodes.}
    \label{tab:IIbSource}
\end{table}
\nocite{*}
\bibliography{references}


\end{document}